\documentclass[%
 reprint,
 amsmath,amssymb,
 aps,
]{revtex4-2}

\usepackage{graphicx}
\usepackage{dcolumn}
\usepackage{bm}
\usepackage{color}
\usepackage{hyperref}
\usepackage{array}
\usepackage{multirow}

\begin{document}

\preprint{APS/123-QED}

\title{High-speed 2D and 3D mid-IR imaging with an InGaAs camera}

\author{Eric O. Potma}
 \email{epotma@uci.edu}
\author{Dave Knez}%
\affiliation{%
Department of Chemistry, University of California, Irvine, CA 92697
}%

\author{Martin Ettenberg}
\author{Matthew Wizeman}
\author{Hai Nguyen}
\author{Tom Sudol}
\affiliation{
Princeton Infrared Technologies, Momouth Jct., NJ 08852, US
}%

\author{Dmitry A. Fishman}
 \email{dmitryf@uci.edu}
\affiliation{%
Department of Chemistry, University of California, Irvine, CA 92697
}%

\date{\today}

\begin{abstract}
Recent work on mid-infrared (MIR) detection through the process of non-degenerate two-photon absorption (NTA) in semiconducting materials has shown that wide-field MIR imaging can be achieved with standard Si cameras. While this approach enables MIR imaging at high pixel densities, the low nonlinear absorption coefficient of Si prevents fast NTA-based imaging at lower illumination doses. Here we overcome this limitation by using InGaAs as the photosensor. Taking advantage of the much higher nonlinear absorption coefficient of this direct bandgap semiconductor, we demonstrate high-speed MIR imaging up to 500 fps with under 1 ms exposure per frame, enabling 2D or 3D mapping without pre- or post-processing of the image.
\end{abstract}

\maketitle


\section{Introduction}
The process of non-degenerate two-photon absorption (NTA) forms an attractive strategy for the detection of low energy photons with wide bandgap semiconductors.\cite{Fishman2011,Cirloganu2011} In NTA the energy needed for the generation of charge carriers in the semiconductor is determined by the sum of the energies of a long wavelength signal photon and a shorter wavelength gate photon. In particular, NTA has made it possible to detect signals in the mid-infrared (MIR) wavelength range, which roughly spans $3~\mu\rm{m}–12~\mu\rm{m}$, with Si-based detectors.\cite{Xu2019,Cox2019,Fang2020} When applied to imaging, detecting MIR light with Si detector technologies offers several advantages compared to the use of low bandgap MIR cameras, including much lower thermal noise and significantly higher pixel densities. For instance, NTA-based imaging with a Si CCD camera has enabled 4~Mpx MIR mapping with 100~ms exposure times.\cite{Knez2020} In addition, using a femtosecond gate pulse, axial optical slicing of the 3D MIR image can be achieved, allowing tomographic mapping with contrast based on the sample’s MIR spectroscopic transitions.\cite{Potma2021} 

NTA-based imaging with Si cameras offers a promising route for MIR mapping at high pixel densities. However, Si is an indirect semiconductor, and its nonlinear absorption coefficient is unfavorably low compared to the ones of direct bandgap materials. Another limitation of Si is its shallow (linear) absorption edge, which translates into a spectral response of the camera that displays a tail on the low energy side, extending from 900~nm to well over 1100~nm. The shallow absorption edge profile limits flexible tuning of gate pulse energies due to one-photon absorption, which renders Si detectors incompatible for NTA-based MIR detection at energies below $\sim900~\rm{cm}^{-1}$. The Urbach tail of direct bandgap semiconductors, on the other hand, generally displays a much steeper profile, and would thus enable NTA detection over a more extended MIR tuning range.

In this work we push the efficiency of NTA-based MIR imaging by selecting a detector based on a direct bandgap semiconducting material. A careful examination of two-photon absorption efficiencies as well as the practical tuning range for MIR detection identifies InGaAs as an ideal candidate for NTA applications. Using an InGaAs camera, we achieve MIR imaging with frame rates that are two orders of magnitude faster than previously shown for Si CCD cameras. We show high-speed 2D and 3D imaging with 1Mpx frames at 100~fps and 40~kpx frames at 500~fps, using exposure times as low as $60~\mu\rm{s}$ per frame. We demonstrate that these new imaging capabilities enable direct \emph{in situ} detection of several mechanical and physio-chemical processes.
\begin{figure*}[ht]
    \centering
    \includegraphics[width=\textwidth]{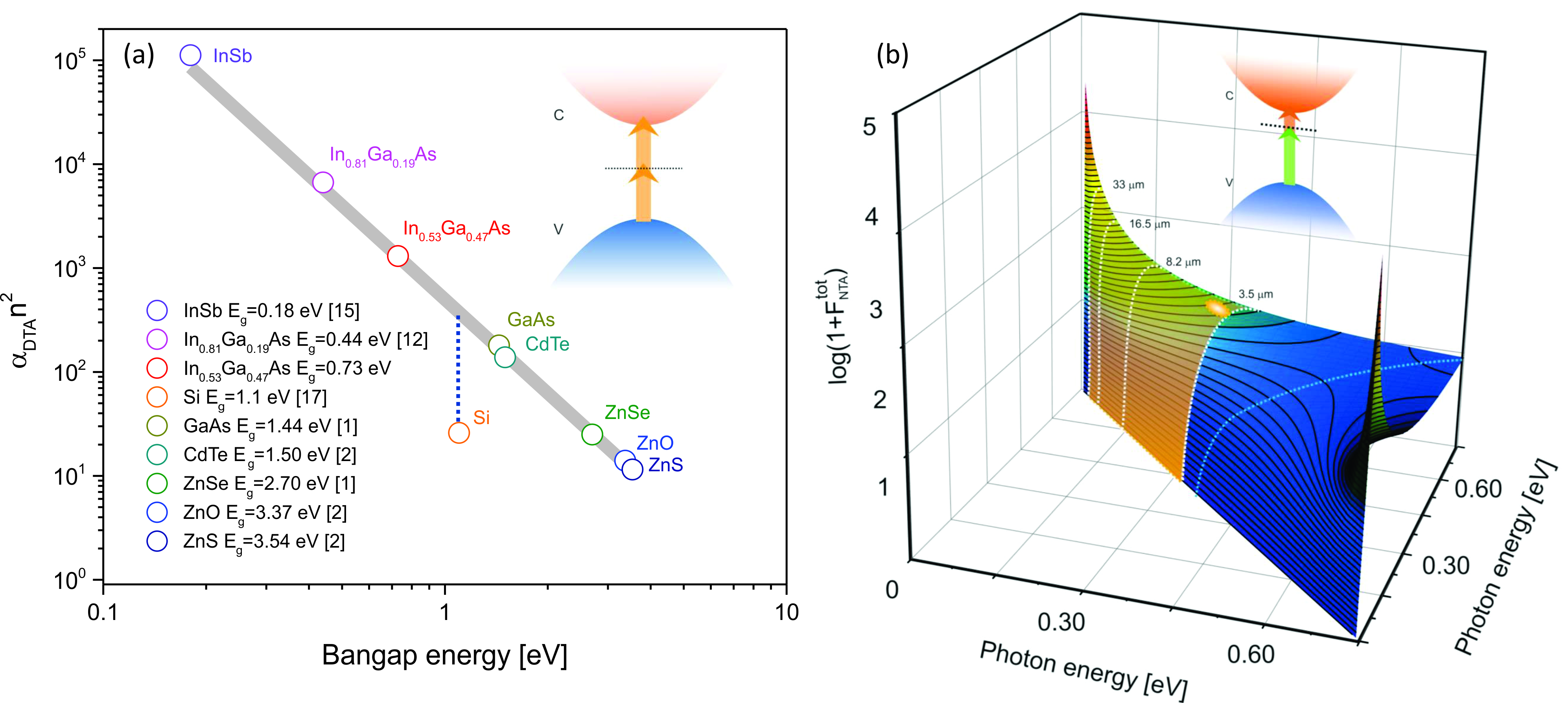}
    \caption{ (a) Scaling rule of degenerate two-photon absorption coefficient as a function of semiconductor bandgap. (b) $F^{tot}_{\rm{NTA}}$ function versus two photon energies calculated for lattice matched InGaAs. The light blue curve represents the case of DTA.}
    \label{fig:fig1}
\end{figure*} 

\section{II. Semiconductors nonlinearity scaling rule}
In the case of linear absorption in semiconducting materials, where one photon excites an electron from the valence to the conduction band, the carrier population in the conduction band scales linearly with incoming optical power. In two-photon absorption, two photons are required to produce a similar transition if the sum of their energies exceeds the bandgap. For degenerate two-photon absorption (DTA), the carrier population scales quadratically with incident optical power. To estimate the efficiency of such process, an elegant quantum-mechanical model based on Keldysh theory has been developed in the 1980s.\cite{Jones1977,Keldysh1965,Wherrett1984} Using a two-band model and a second-order perturbation approach, the DTA coefficient can be expressed as follows:\cite{VanStryland1988}
\begin{eqnarray}
    \alpha_{\rm{DTA}}&=&K\sqrt{E_p}\frac{F_{\rm{DTA}}(x)}{n^2E^3_g}\label{eq:alpha_DTA}\\
    F_{\rm{DTA}}(x)&=&\frac{(2x-1)^{3/2}}{2^5x^5},\;\;\;\;\;\;x=\frac{\hbar\omega}{E_g}
\end{eqnarray}
where $E_p$ is the Kane energy, $K$ is a material independent constant, $n$ is the linear refractive index of the material and $E_g$ is the semiconductor bandgap. This model predicted and later experimentally confirmed the DTA response of a wide variety of narrow and wide direct bandgap semiconductors. Following the logic of equation (\ref{eq:alpha_DTA}), $\alpha_{\rm{DTA}} n^2$ scales inversely with $E_g^3$. Figure 1a represents this scaling rule, overlaid with experimental data acquired over several decades.\cite{Fishman2011,Cirloganu2011,Piccardo2018,Pattanaik2015,Cirloganu2010,Johnston1980} This graph makes it clear that the results for Si do not follow the general trend. The reason why a two-band model unsuccessfully predicts the transition probability for Si is because it does not capture the effect of the additional phonon mode needed to mediate the transition in momentum space. Several approaches utilizing different transition pathways have been proposed, among which are the ``forbidden-forbidden''~\cite{Dinu2003}, ``allowed-forbidden'' and ``allowed-allowed'' models.\cite{Bristow2007,Garcia2006} However, a more generalized theory is yet to be developed.

\begin{figure*}[ht]
    \centering
    \includegraphics[width=\textwidth]{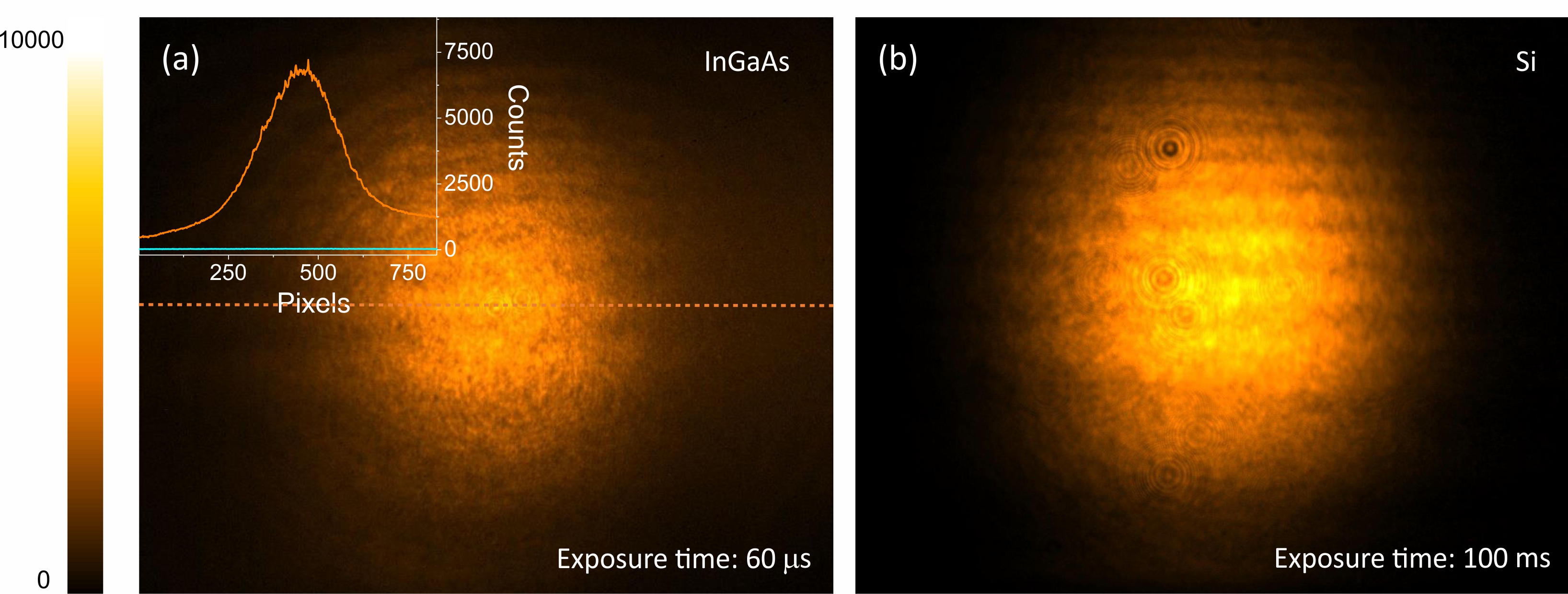}
    \caption{(a) 4250~nm ($\sim2350~\rm{cm}^{-1}$) single shot detection by InGaAs camera using $60~\mu\rm{s}$ exposure time. The MIR average power is $130~\mu\rm{W}$. Inset shows the spatial cross-section of the beam profile on the camera chip (orange curve) when both MIR and gate pulse are present. The light blue curve shows the DTA background when only the 1900~nm gate beam is present. (b) 3840~nm ($\sim2600~\rm{cm}^{-1}$) pulse train detection on a Si CCD camera using 100~ms exposure time. The average power is 1~mW.}
    \label{fig:fig2}
\end{figure*}

The DTA theory can be expanded for the case of non-degenerate two-photon absorption, where the nonlinear absorption coefficient is now expressed as:\cite{SheikBahae1992,Hutchings1992,SheikBahae1991} 
\begin{eqnarray}
\alpha_{\rm{NTA}}(x_1,x_2)=K\sqrt{E_p}\frac{F_{\rm{DTA}}(x_1,x_2)}{n_1n_2E^3_g}\label{eq:alpha_NTA}\\
F_{\rm{NTA}}(x_1,x_2)=\frac{(x_1+X-2-1)^{3/2}}{2^7x_1^3x_2^4}(x_1+x_2)^2\label{eq:F_NTA}\\
x_1=\frac{\hbar\omega_1}{E_g},\;\;x_2=\frac{\hbar\omega_2}{E_g}
\end{eqnarray}
here $n_1$ and $n_2$ are the refractive indices of the material at the respective wavelengths of the incident photons. It is useful to relate the nonlinear absorption coefficient defined in (\ref{eq:alpha_NTA}) to the actual charge carriers generated in the semiconductor, as the number of photo-excited conduction electrons is directly proportional to the signal registered in NTA detection applications. The change in the number of photo-excited electrons $N_c$ in the conduction band with respect to a change in the propagation distance $z$ in the material depends on the loss of photon numbers $N_1$ and $N_2$, and can be expressed as:
\begin{equation}
    \frac{dN_c}{dz}=-\frac{1}{2}\left\{ \frac{dN_1}{dz}+ \frac{dN_2}{dz}\right\}\label{eq:carriers_diff}
\end{equation}
with
\begin{eqnarray}
 \frac{dN_1}{dz}&=&-2\alpha_{\rm{NTA}}(x_1,x_2)\frac{\hbar\omega_2}{\tau\pi\omega^2_0}N_1N_2\label{eq:N1}\\
 \frac{dN_1}{dz}&=&-2\alpha_{\rm{NTA}}(x_2,x_1)\frac{\hbar\omega_1}{\tau\pi\omega^2_0}N_1N_2\label{eq:N2}
\end{eqnarray}
where $\hbar\omega_1$ and $\hbar\omega_2$ are the photon energies, expressed in Joules, $\tau$ is the interaction time between the two beams and $\pi\omega^2_0$ is the illumination area. From equations (\ref{eq:carriers_diff})-(\ref{eq:N2}), it follows that the change in number of electrons in the conduction band can be written as:
\begin{equation}
\frac{dN_c}{dz}=\frac{K\sqrt{E_p}}{n_1n_2\tau\pi\omega_0}\frac{N_1N_2}{E^2_g}F^{tot}_{\rm{NTA}}
\end{equation}
where
\begin{eqnarray}
F^{tot}_{\rm{NTA}}&=&x_2F_{\rm{NTA}}(x_1,x_2)+x_1F_{\rm{NTA}}(x_2,x_1)\nonumber\\
&=&\frac{(x_1+X-2-1)^{3/2}}{2^6x_1^3x_2^3}(x_1+x_2)^2\label{eq:F_tot}
\end{eqnarray}
It is interesting to note that the detected NTA signal, which follows from (3) is inversely proportional to $E_g^2$, in contrast to the $E_g^3$ scaling of individual nonlinear coefficients $\alpha_{\rm{NTA}}(x_1,x_2)$ and $\alpha_{\rm{NTA}}(x_2,x_1)$. See the {\href{https://www.chem.uci.edu/~dmitryf/images/Supplementary Information InGaAs.pdf}{{\color{blue} Supplementary Information}}} for a more detailed derivation of equation (3).

The function $F^{tot}_{\rm{NTA}}$ is symmetric in $\hbar\omega_1$ and $\hbar\omega_2$, and is plotted for the case of InGaAs in Figure \ref{fig:fig1}b. Together, Figures  \ref{fig:fig1}a and  \ref{fig:fig1}b highlight a few important advantages of InGaAs for broadband MIR detection through NTA. First, the low bandgap of InGaAs gives rise to strong degenerate nonlinear absorption coefficients, yet $E_g$ of the material is large enough to suppress thermally induced transitions relative to thermal effects seen in InSb and HgCdTe. The degenerate case is schematically depicted by the blue line in Figure 1b with a maximum of $\alpha_{\rm{DTA}}=135$~cm/GW for photon energies of $\hbar\omega=0.7E_g$. Second, the non-degenerate two-photon absorption efficiency increases dramatically relative to $\alpha_{\rm{DTA}}$ when the photon energy ratio disparity $\hbar\omega_1/\hbar\omega_2$ deviates from unity, as predicted by equation (\ref{eq:F_tot}). This enables detection over multiple spectral octaves from $3.5~\mu\rm{m}$ to $>30~\mu\rm{m}$ (Figure  \ref{fig:fig1}b) with $\alpha_{\rm{NTA}}$  approaching tens of MW/cm when $\hbar\omega_{\rm{IR}}$ is tuned to lower energies.

\section{Materials and Methods}
The experimental setup has been described in previous work.\cite{Potma2021} Particular parameters for the current studies are given in {\href{https://www.chem.uci.edu/~dmitryf/images/Supplementary Information InGaAs.pdf}{{\color{blue} Supplementary}}} Table 1. Two femtosecond pulses, a tunable MIR beam and a fixed NIR gate beam at 1900~nm (5263~cm$^{-1}$), are spatially overlapped on the InGaAs camera chip (1280MVCam, Princeton Infrared Technologies, Inc.). The chip is based on lattice matched In0.53Ga0.47As alloy, which exhibits a steep bandgap absorption edge around 1700~nm (0.73~eV) at room temperature. The quantum efficiency and spectral responsivity curves are presented in Supplementary Figures S1 and S2. The camera enables high-speed imaging utilizing the whole chip of $1280\times1024$ pixels ($12~\mu\rm{m}$ pixel pitch) up to 100~fps. Faster voltage readout requires reduction of the region of interest, enabling 500~fps for $200\times200$ pixel frames. It is important to note that the camera chip is protected by a borosilicate window (thickness 2 mm), which significantly attenuates the MIR radiation ($\rm{OD}=1.25$ at 4200~nm, $\rm{OD}>5$ at 5000~nm, see {\href{https://www.chem.uci.edu/~dmitryf/images/Supplementary Information InGaAs.pdf}{{\color{blue} Supplementary}}} Figure S2). The estimated pulse energies are $\sim200$~fJ per pixel for the MIR beam ($130~\mu\rm{W}$ average power) and $\sim50$~fJ per pixel for the 1900~nm gate beam ($20~\mu\rm{W}$ average power). The MIR beam is directed to the sample and scattered light is collected by a 100~mm CaF$_2$ lens, forming an image of the sample onto the camera in a 1:1 fashion. An NTA signal is generated in the InGaAs chip whenever the MIR temporally overlaps with the NIR gate pulse. The current magnification and effective numerical aperture of the imaging lens ($\rm{NA}=0.015$) provides an image with $\sim100~\mu\rm{m}$ resolution, corresponding to $\sim20$ pixels on the camera. Better spatial resolution, or a larger field of view, can be achieved by changing the imaging lens.
\begin{figure*}[ht]
    \centering
    \includegraphics[width=\textwidth]{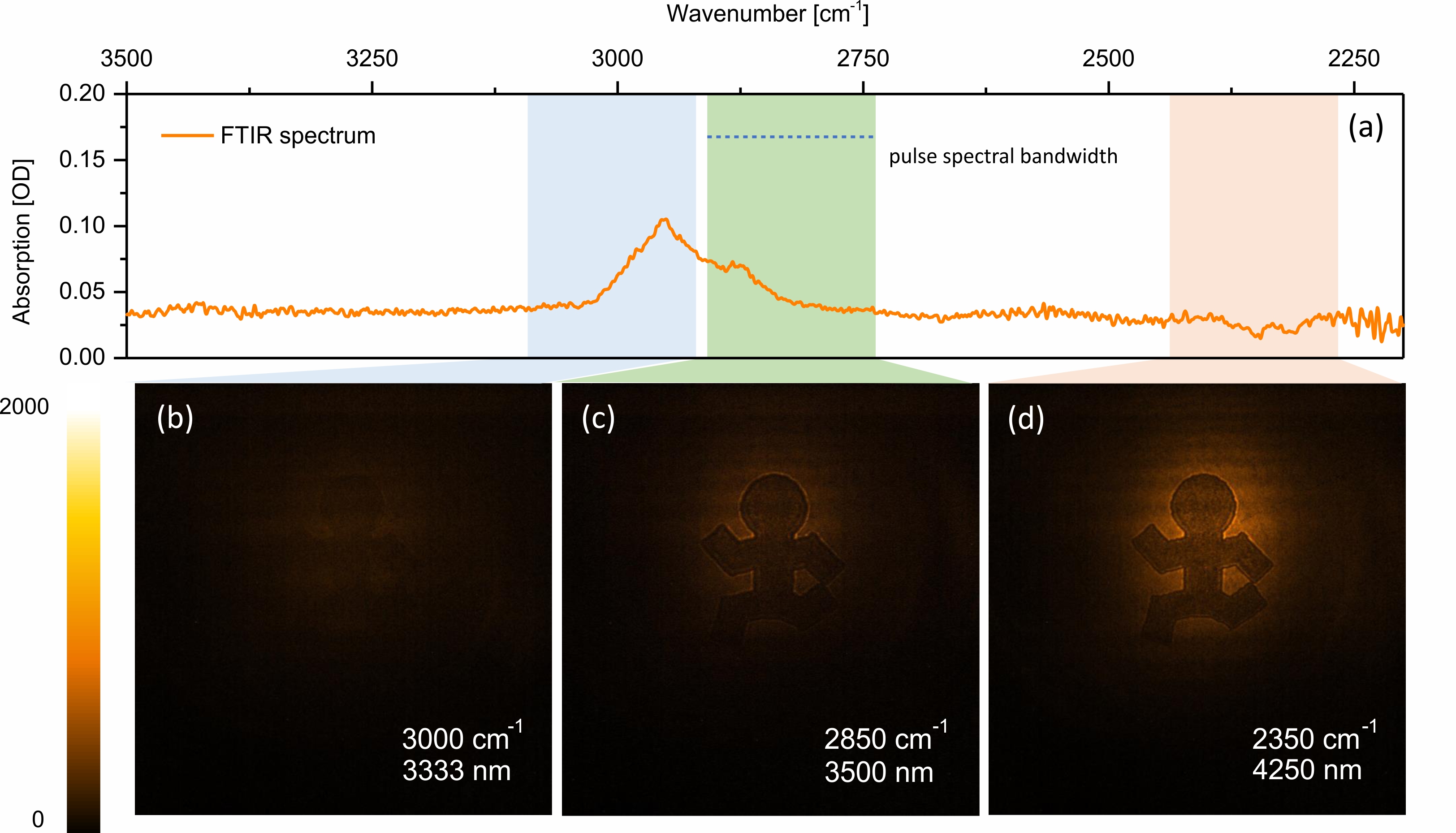}
    \caption{Chemically selective 2D imaging of a cellulose acetate film. A figurine is printed on the film with black ink to generate contrast. If tuned to the resonance of C-H stretching modes, cellulose acetate becomes less transparent, decreasing the contrast between the clear and inked areas of the film. Exposure time 1 ms. See full videos at Table 1, Multimedia view MV1, MV2 and MV3.}
    \label{fig:fig3}
\end{figure*}

\section{Results and Discussion}

\subsection{Short exposure time imaging}
Figure \ref{fig:fig2}a shows an NTA image of the MIR beam at 4250~nm ($\sim2350~\rm{cm}^{-1}$) on the InGaAs chip. The high optical nonlinearity of the InGaAs material enables efficient NTA, allowing an exposure time of only $60~\mu\rm{s}$. As such fast exposure times, light contamination from the surrounding is negligible. At the 1~kHz repetition rate of the light source, the inter-pulse separation is 1~ms, which is much longer than the exposure time. Consequently, the signal shown in Figure \ref{fig:fig2}a is caused by a single MIR/gate pulse pair. Similarly, for exposure times less than 1~ ms, the NTA signal in illuminated frames is derived from single shots that fall within the exposure time window.

The DTA background can be suppressed when the gate pulse power is lowered relative to the pulse power of the MIR beam, allowing background-free imaging experiments. In Figure \ref{fig:fig2}a, the overall signal-to-background ratio is around 20~dB ($10\log(S_{\rm{NTA}}/S_{\rm{DTA}}$) , with signal-to-noise of 33~dB ($10\log(S_{\rm{NTA}}/\sigma_{\rm{DTA}}$) or 66~dB (mean root power, $20\log(S_{\rm{NTA}}/\sigma_{\rm{DTA}}$), where $\sigma_{\rm{DTA}}$ is the standard deviation of the DTA background signal.

For comparison, Figure \ref{fig:fig2}b shows an NTA image of the same MIR beam collected with a Si CCD camera (Clara, Andor Technologies). The average powers of the MIR and gate beams, as well as the exposure time, are adjusted to achieve roughly the same NTA signal level on the Si camera. Accounting for the differences in illumination conditions, optical properties of protective windows, wavelength-dependent quantum efficiencies of the respective sensors and gain settings we observe that the InGaAs camera exhibits a ~110 times higher detection efficiency than its Si counterpart. These observations are in good agreement with equation (\ref{eq:alpha_NTA}), which predicts nonlinear coefficients of 355~cm/GW for InGaAs and 3~cm/GW for Si for the photon energies used in the experiment.

\subsection{Chemically selective MIR videography}
The short exposure time enabled by the higher $\alpha_{\rm{NTA}}$ of InGaAs opens up the possibility of high-speed MIR videography. Table 1 contains snapshots from MIR videos of moving targets, recorded at 30~fps with 1~ms exposure time per frame. Under these conditions, in each frame the chip is illuminated on average by a single MIR/gate pulse pair. Among the examples is a structure printed on a $\sim110~\mu\rm{m}$ thick cellulose acetate film, shown in more detail in Figure \ref{fig:fig3}. One of the advantages of MIR imaging is the capability to generate contrast based on spectroscopic transitions in the material, underlined by the FTIR spectrum of cellulose acetate depicted in Figure \ref{fig:fig3}a. The absorption feature in the relevant spectral range corresponds to the C-H stretching vibrational mode. In Figure \ref{fig:fig3}b, the MIR beam is tuned to 3000~cm$^{-1}$ where absorption is significant, reducing the transmission of the MIR light through the material, resulting in poor image contrast. When the MIR beam is tuned off the vibrational resonance, light absorption decreases, enabling more light to pass through the clear part of the film, as illustrated by Figures \ref{fig:fig3}c and \ref{fig:fig3}d. 

\begin{table}[t]
\centering
\caption{Representative compressed videos of moving targets and live organisms. Frame rate 30 fps with exposure time of 1 ms per frame.\label{tab:Table1}}
\begin{tabular}{ |>{\centering\arraybackslash} p{3cm}| >{\centering\arraybackslash} m{3cm}| >{\centering\arraybackslash} m{2cm}| }
 \hline
video snapshot & description & link \\ 
 \hline
{\includegraphics[width=3cm]{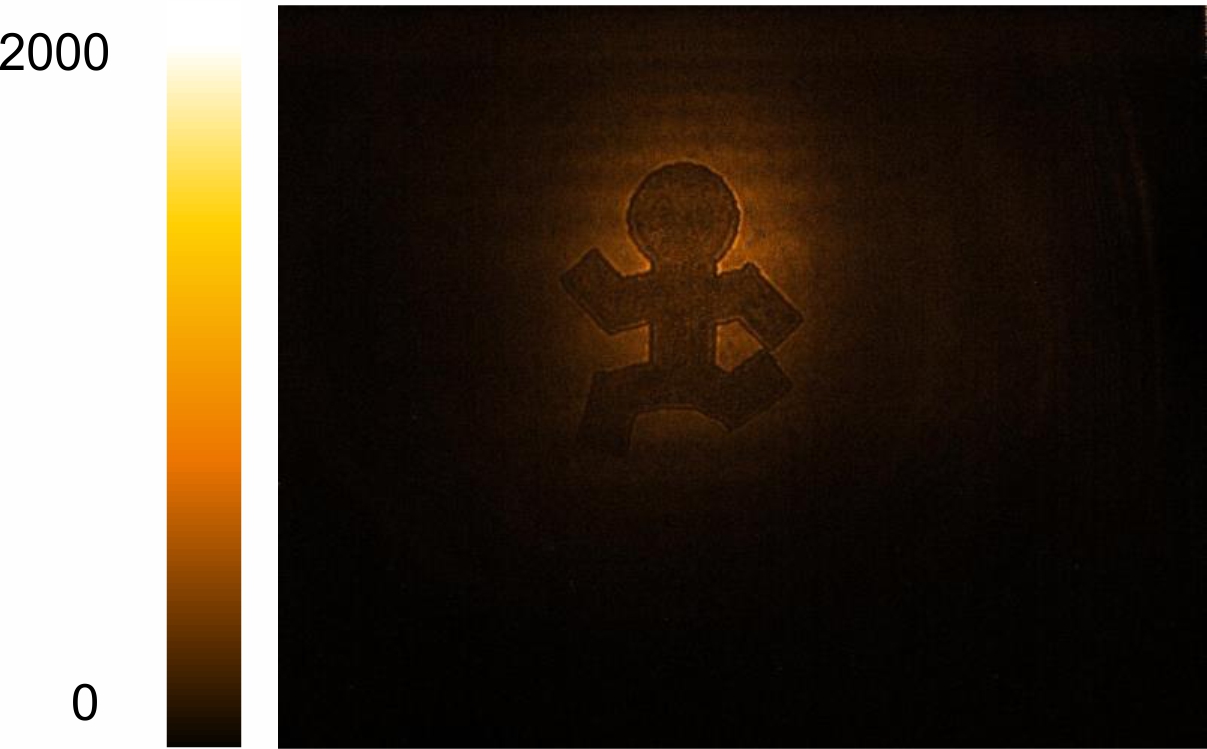}}
 &{\small{Printed target on cellulose acetate film. Movie complementary to Figure \ref{fig:fig3}}.}
 & {\href{https://www.chem.uci.edu/~dmitryf/images/Geronimo2600cm-1.mp4}{{\color{blue} 2600~cm$^{-1}$}} \href{https://www.chem.uci.edu/~dmitryf/images/Geronimo2850cm-1.mp4}{{\color{blue} 2850~cm$^{-1}$}}
 \href{https://www.chem.uci.edu/~dmitryf/images/Geronimo3000cm-1.mp4}{{\color{blue} 3000~cm$^{-1}$}}}\\
\hline
\includegraphics[width=3cm]{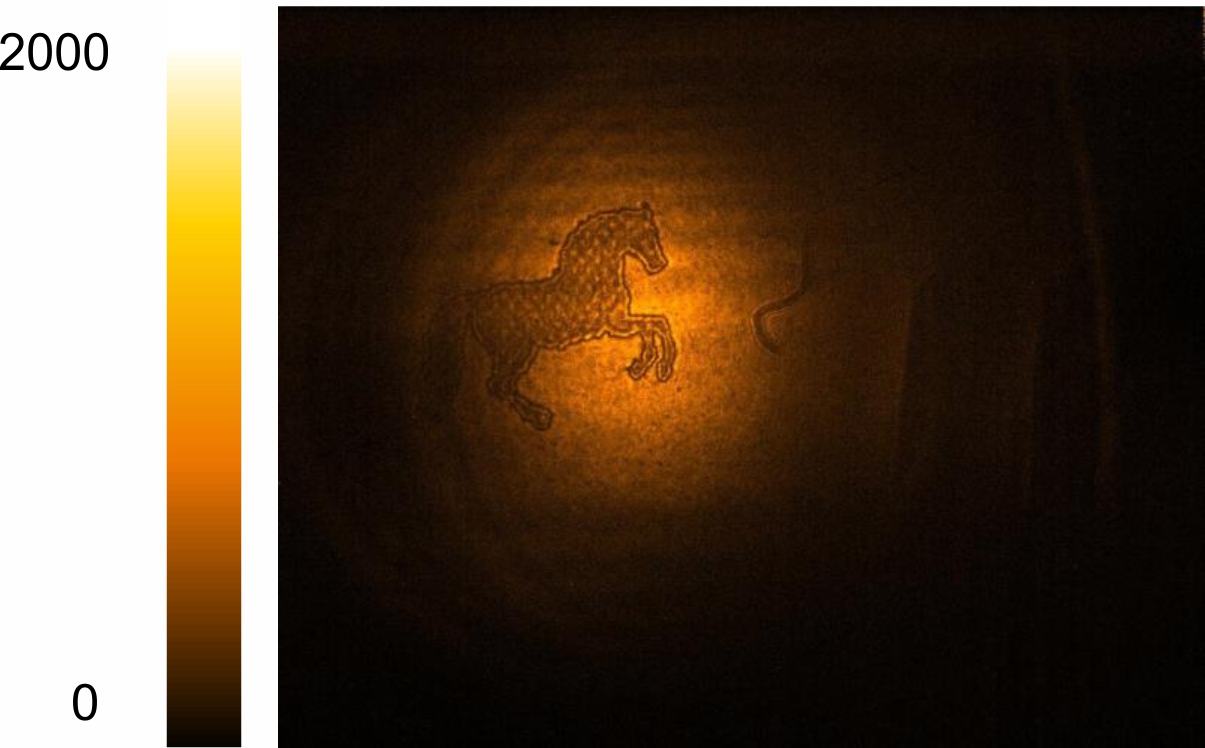}
 &\small{Printed target on cellulose acetate film. MIR at 4250~nm ($\sim2350~\rm{cm}^{-1}$)}.
 & {\href{https://www.chem.uci.edu/~dmitryf/images/Bucephalus.mp4}{{\color{blue} MV4}}}\\
\hline
\includegraphics[width=3cm]{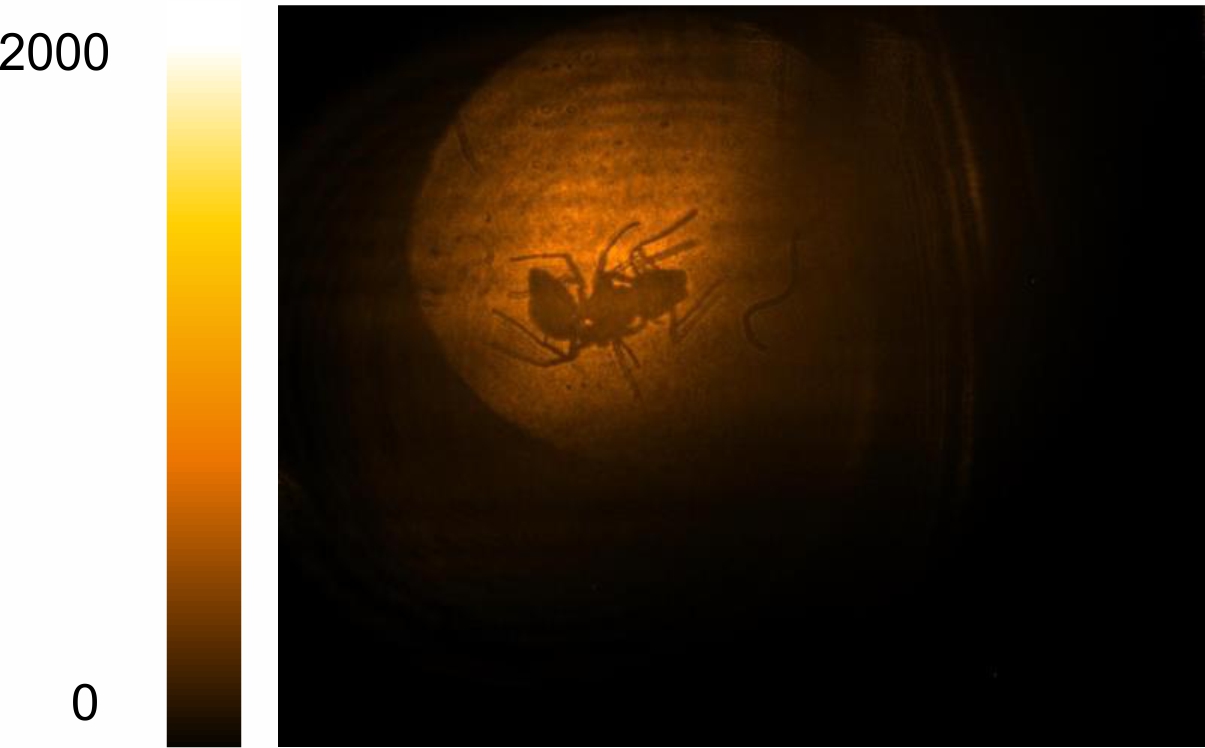}
 &\small{Live recording of Tetramorium caespitum (pavement ant). MIR at 4250~nm ($\sim2350~\rm{cm}^{-1}$)}.
 & {\href{https://www.chem.uci.edu/~dmitryf/images/Hector.mp4}{{\color{blue} MV5}}}\\
 \hline
\end{tabular}
\end{table}

\begin{figure}[hb]
    \centering
    \includegraphics[width=8cm]{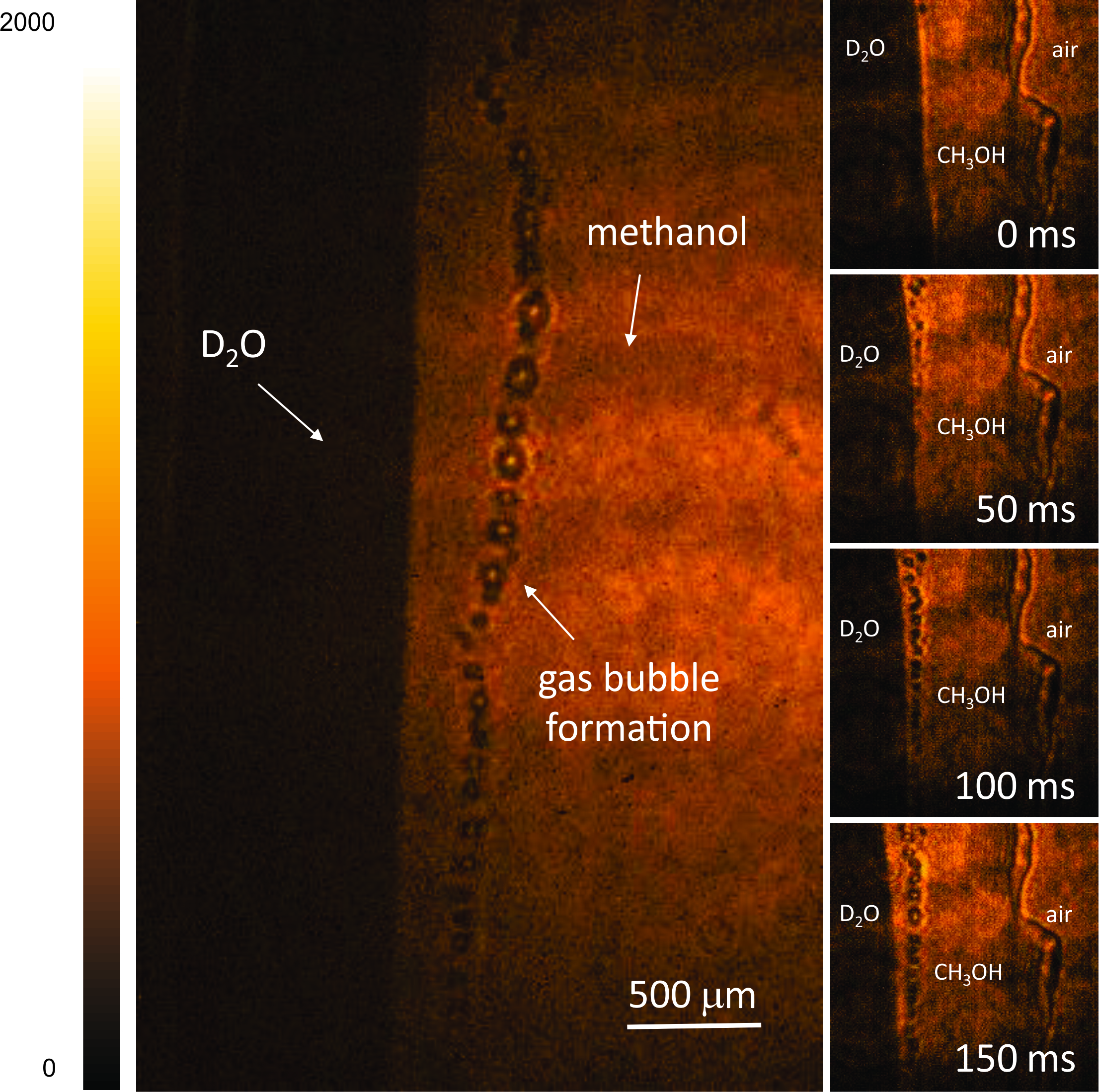}
    \caption{Snapshots of chemically sensitive 2D time-lapse imaging of the early dynamics of a methanol-D$_2$O mixture (full videos in Table 2). D$_2$O appears much darker due to the D-O stretching absorption near $2600~\rm{cm}^{-1}$ (3850~nm), while methanol remains nearly transparent. Bubbles are forming in the first 150~ms. Exposure time 1~ms.}
    \label{fig:fig4}
\end{figure}

Table 2 contains thumbnails of video recordings of several mixing dynamics experiments based on methanol and D$_2$O. In these mixing experiments, the active area is visualized in the transmission geometry at acquisition rates of up to 500~fps. In the visible range both liquids are transparent, hence indistinguishable. However, in the MIR spectral region, due to the distinct absorption band of D-O stretching vibrations, D$_2$O appears much darker at 2600~cm$^{-1}$ (3850~nm), while methanol remains nearly transparent, as shown in Figure \ref{fig:fig4}. Since mixing and hydrogen bonding between methanol and D$_2$O molecules is energetically favorable, the process is slightly exothermic.\cite{Boyne1965,Peeters1993} In the early mixing dynamics the formation of sub-mm bubbles is observed~\cite{Rage2020}, which can possibly be attributed to the mechanism of solvated gas release (N$_2$, O$_2$, CO$_2$) induced by the local increase of temperature near the interfacial regions.\cite{Messel1966,Millare2018} 
\begin{table}[ht]
\centering
\caption{Time lapse imaging of mixing process of methanol and D$_2$O, with frame rates of 30~fps, 100~fps and 500~fps, and an exposure time of 1~ms per frame.\label{tab:Table2}}
\begin{tabular}{ |>{\centering\arraybackslash} p{3cm}| >{\centering\arraybackslash} m{3cm}| >{\centering\arraybackslash} m{2cm}| }
 \hline
video snapshot & description & link \\ 
 \hline
{\includegraphics[width=3cm]{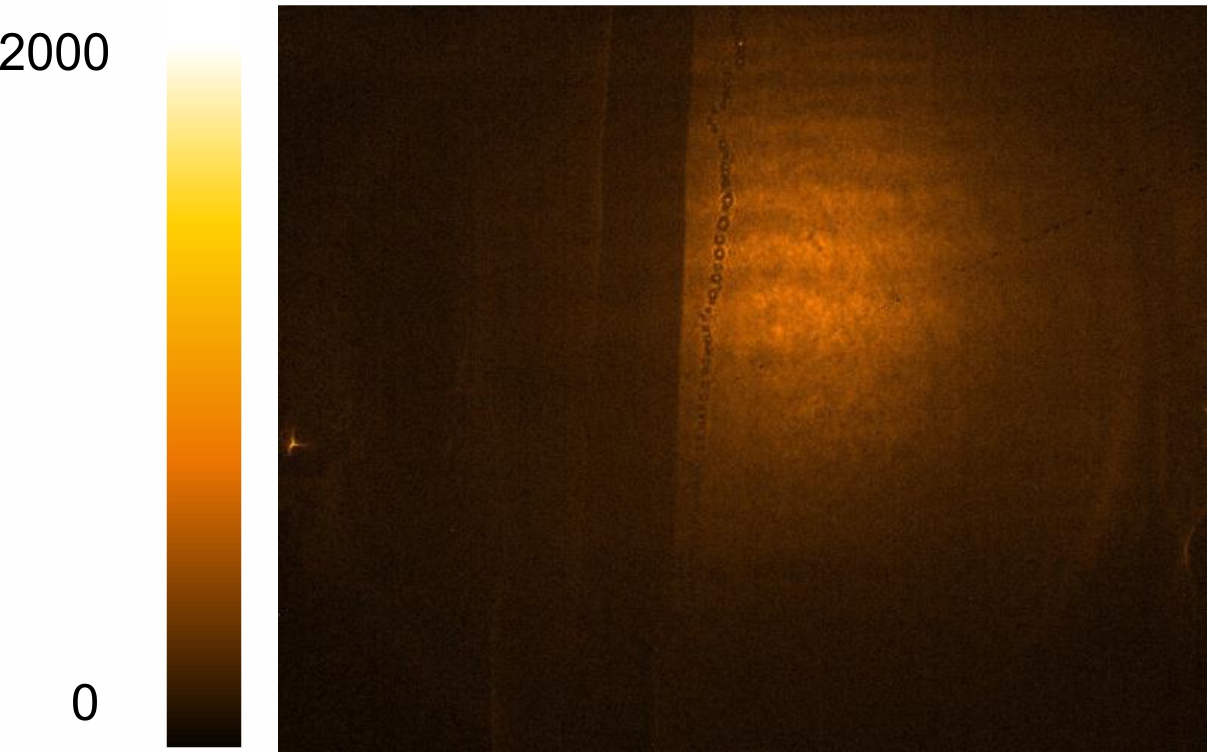}}
 &{\small{Methanol and D$_2$O mixture on a microscope cover slip. MIR at 2600 cm$^{-1}$.}}
 & {\href{https://www.chem.uci.edu/~dmitryf/images/Liquids30fps.mp4}{{\color{blue} MV6}}}\\
\hline
\includegraphics[width=3cm]{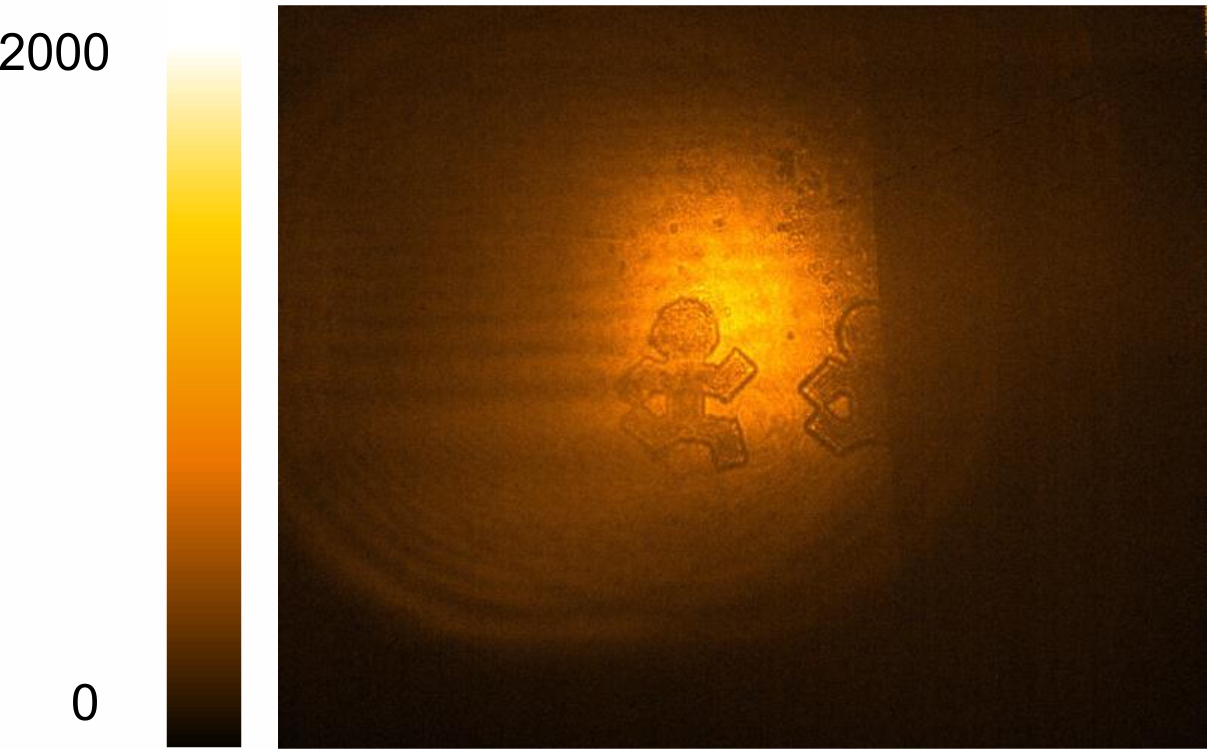}
 &\small{Methanol and D$_2$O mixture on a cellulose acetate film. MIR at 2600 cm$^{-1}$.}
 & {\href{https://www.chem.uci.edu/~dmitryf/images/Liquids100fps.mp4}{{\color{blue} MV7}}}\\
\hline
\includegraphics[width=3cm]{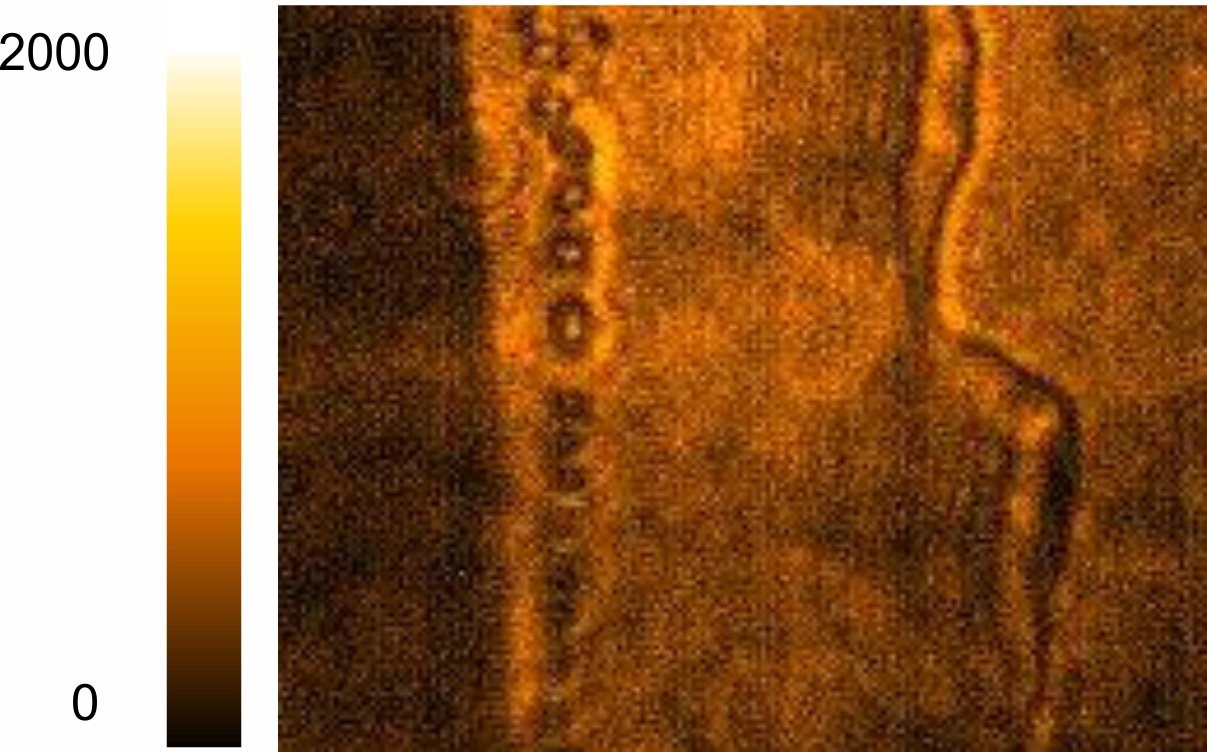}
 &\small{Solvated gas bubble formation at 500~fps. MIR at 2600 cm$^{-1}$. Region of interest is $200\times200$~px.}
 & {\href{https://www.chem.uci.edu/~dmitryf/images/Liquids500fps.mp4}{{\color{blue} MV8}}}\\
 \hline
\end{tabular}
\end{table}

\subsection{High-speed 3D imaging}
\begin{figure*}[ht]
    \centering
    \includegraphics[width=\textwidth]{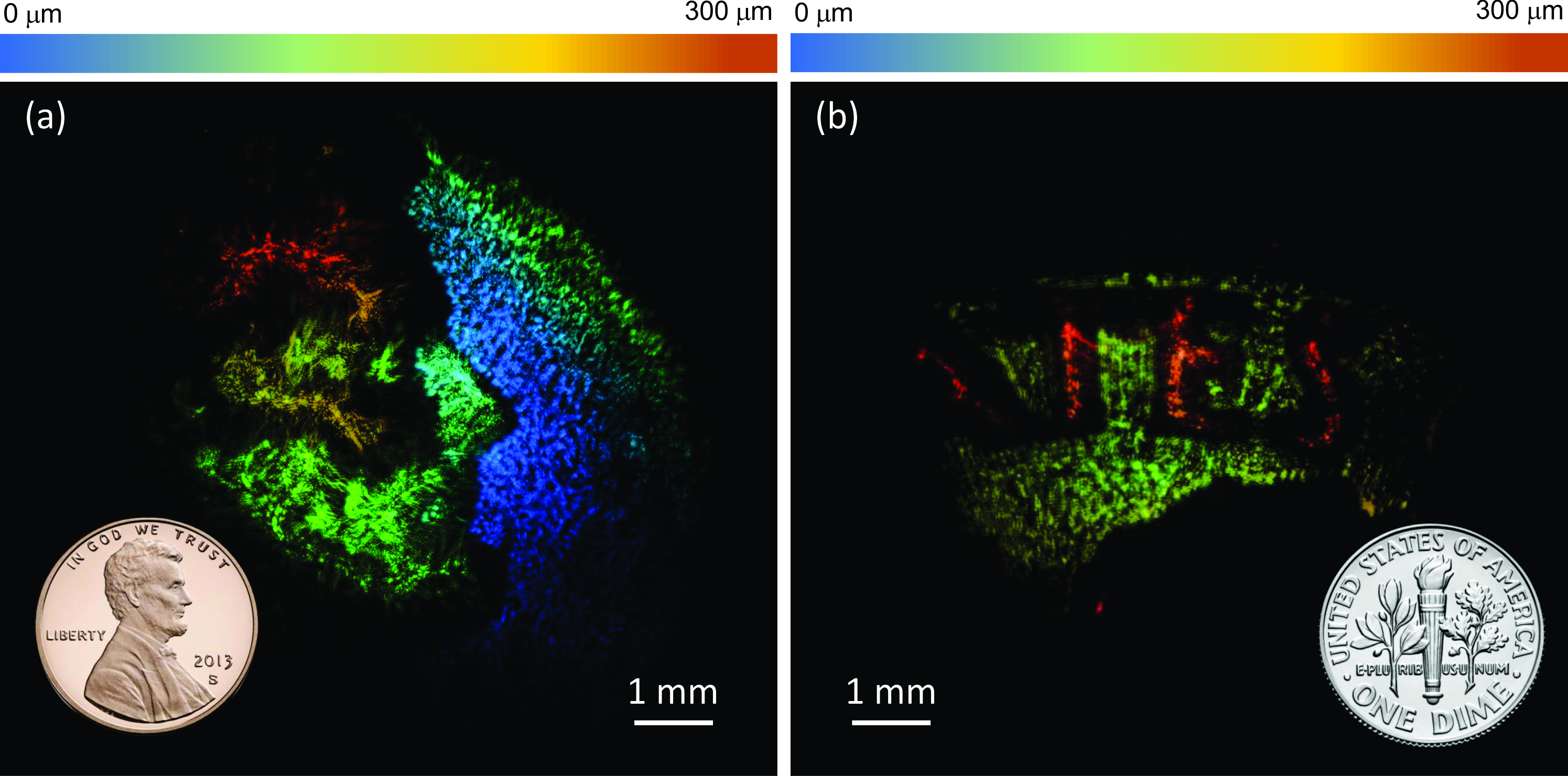}
    \caption{Tomographic images of US penny (a) and dime (b) scanned at 4250~nm ($2350~\rm{cm}^{-1}$). Total acquisition time for each volumetric scan is 20~ms.}
    \label{fig:fig5}
\end{figure*}

The use of fs pulses in NTA imaging enables axial optical slicing through temporal gating by the NIR pulse, thus making 3D imaging possible[7]. In this tomographic MIR imaging approach, the depth scan is realized by adjusting the time delay between the MIR and gate pulses at the camera chip. Since the time delay can be scanned rapidly, fs-NTA allows for rapid acquisition of 3D image stacks. When combined with the increase in acquisition speed afforded by the InGaAs camera, truly high-speed 3D imaging in the MIR is within reach. In Figure \ref{fig:fig5}, we apply this principle to visualize features on the US penny and dime. The axial resolution is determined by the short coherence length of the femtosecond pulses, which in the current work is $\sim10~\mu\rm{m}$. Each 2D image frame is acquired in 1ms, and with 20 images in the 3D data stack the effective acquisition time is 20~ms. Note that such volumetric imaging rates are more than 50 times higher than previously reported[7], emphasizing the drastic improvement of imaging performance with InGaAs cameras. 

\section{Conclusion}
In this work we have significantly advanced the speed of NTA-based MIR imaging by using an InGaAs camera. Compared to Si detectors, the high nonlinear absorption coefficient of InGaAs sensors increases the efficiency of NTA-induced photocurrents by over two orders of magnitude. We have shown that the higher NTA efficiency readily improves the imaging speed by a factor of at least 100 times. At such imaging rates, 2D and 3D visualization of fast dynamic processes that previously were inaccessible with MIR imaging becomes possible. In particular, using the improved imaging capabilities, we have successfully resolved liquid-liquid mixing dynamics of methanol and D$_2$O on the millisecond timescale, revealing the formation of gas bubbles within a ~100ms time window. In addition, we have demonstrated the acquisition of 3D images within tens of ms, acquisition speeds that are unprecedented for tomographic mapping in the MIR.

The faster NTA-based MIR imaging capabilities have implications beyond fast videography or 3D imaging. Using chirped broadband MIR pulses, a rapid time-delay scan of the gate pulse would allow for fast spectral sweeping, resulting in the acquisition of hyperspectral data stacks. We anticipate that such data stacks can be recorded at rates exceeding one hyperspectral image per second, hence permitting spectrally-resolved imaging of sub-second processes.  Whereas the current work uses fs pulses, the high NTA efficiencies demonstrated here open the door for utilization of lower irradiance sources, including commercially available quantum cascade lasers (QCLs). In addition, lower doping levels of indium could shift the In1-xGaxAs absorption edge to higher energies according to $E_g=0.36+0.63x+0.47x^2$.\cite{Goetz1983} This would render the InGaAs detector suitable for NTA with gate photons derived from Er-doped telecom lasers. Such developments will bring the NTA-based imaging technology into the realm of compact and affordable light sources.

\begin{acknowledgments}
We thank the National Institutes of Health, grant GM R21-GM141774 and Air Force STTR AFX20C-TCS01/AFWERX Program under contract F4FBEQ1019A0DP.
\end{acknowledgments}

\appendix


\bibliography{library}

\end{document}